# Pauli String Partitioning Algorithm with the Ising Model for Simultaneous Measurements

*Tomochika Kurita[1,*] Mikio Morita[2] Hirotaka Oshima[1] and Shintaro Sato[1]*

1. Quantum Laboratory, Fujitsu Research, Fujitsu Limited. 10-1 Morinosato-wakamiya, Atsugi, Kanagawa, Japan 243-0197
2. Quantum Laboratory, Fujitsu Research, Fujitsu Limited. 4-1-1 Kami-odanaka, Nakahara-ku, Kawasaki, Kanagawa, Japan 211-8588

* kurita.tomo@fujitsu.com

Abstract:
We propose an efficient algorithm for partitioning Pauli strings into subgroups, which can be simultaneously measured in a single quantum circuit. Our partitioning algorithm drastically reduces the total number of measurements in a variational quantum eigensolver for a quantum chemistry, one of the most promising applications of quantum computing. The algorithm is based on the Ising model optimization problem, which can be quickly solved using an Ising machine. We develop an algorithm that is applicable to problems with sizes larger than the maximum number of variables that an Ising machine can handle ($n_{\text{bit}}$) through its iterative use. The algorithm has much better time complexity and solution optimality than other algorithms such as Boppana–Halldórsson algorithm and Bron–Kerbosch algorithm, making it useful for the quick and effective reduction of the number of quantum circuits required for measuring the expectation values of multiple Pauli strings. We investigate the performance of the algorithm using the second-generation Digital Annealer, a high-performance Ising hardware, for up to 65,535 Pauli strings using Hamiltonians of molecules and the full tomography of quantum states. We demonstrate that partitioning problems for quantum chemical calculations can be solved with a time complexity of $O(N)$ for $N \leq n_{\text{bit}}$ and $O(N^2)$ for $N > n_{\text{bit}}$ for the worst case, where $N$ denotes the number of candidate Pauli strings and $n_{\text{bit}} = 8,192$ for the second-generation Digital Annealer used in this study. The reduction factor, which is the number of Pauli strings divided by the number of obtained partitions, can be 200 at maximum.

1. Introduction

Quantum computing has the potential to outperform classical computing in computational time [1]. In particular, among its several practical targets, there has been a major advancement in the areas of quantum chemistry [2]. In the current noisy intermediate-scale quantum computing [3], variational quantum eigensolver (VQE) algorithms are extensively studied for quantum chemistry to calculate ground- and excited-state energies of chemicals [4,5], including small molecules [6,7], catalysts, and battery materials [8,9].

VQE algorithms are designed to solve the Schrödinger equation,
$$H\psi = E\psi, \qquad (1)$$
using variational methods. To solve it using a quantum computer, the Hamiltonian $H$ and wavefunction $\psi$ are mapped to $\hat{H}$ and $\hat{\psi}$, respectively, through a second quantization:

$$\widehat{H}\widehat{\psi} = E\widehat{\psi}, \tag{2}$$

where $\widehat{\psi}$ can be obtained by a quantum computer. To calculate the ground-state energy, a parameterized quantum state $\widehat{\psi}(\boldsymbol{\theta})$ is created using a quantum computer. The parameters $\boldsymbol{\theta} = \{\theta_1, \theta_2, \cdots\}$ are iteratively optimized using a classical computer to minimize the expectation value of the given Hamiltonian $\widehat{H}$:

$$E = \min_{\boldsymbol{\theta}} \frac{\langle\widehat{\psi}(\boldsymbol{\theta})|\widehat{H}|\widehat{\psi}(\boldsymbol{\theta})\rangle}{\langle\widehat{\psi}(\boldsymbol{\theta})|\widehat{\psi}(\boldsymbol{\theta})\rangle}. \tag{3}$$

To estimate the expectation value $\langle\widehat{\psi}(\boldsymbol{\theta})|\widehat{H}|\widehat{\psi}(\boldsymbol{\theta})\rangle$ using a quantum computer, the Hamiltonian is decomposed into some Pauli strings:

$$\widehat{H} = \sum_i \lambda_i P_i, \tag{4}$$

where $P_i$ denotes the $i$-th Pauli string and $\lambda_i \in \mathbb{R}$ denotes the corresponding weight of $P_i$.

For the VQE algorithms, the number of Pauli strings scales as $O(n^4)$, where $n$ denotes the number of qubits assigned to spin orbitals of a target molecule by a one-to-one correspondence, because the corresponding Hamiltonians only contain two-body interactions. A quite large $n$ value (i.e. a large number of spin orbitals) leads to a large number of required measurements, which limits the ability of quantum computation in general. In principle, to obtain multidimensional information of a target quantum state, we have to prepare its several copies and obtain each one-dimensional information repeatedly through a basis-changing operation prior to each measurement. Furthermore, we need to measure a target quantum state multiple times for each of its one-dimensional information to obtain the expectation value with some desired precision because each measurement results in the projection of the state onto a measurement basis and we need its average value. This estimation process nature of the expectation value makes the overall algorithm time consuming; this condition is especially true when the expectation values of Hamiltonians are subject to classical optimization as in the case of VQEs.

Some methods have been proposed to suppress an increase in the number of measurements [10-18]. One of the main methods is the partitioning method, where Pauli strings are partitioned so that their expectation values can be measured simultaneously [10-14]. In this method, a group of Pauli strings is divided into subgroups, and all the components of each subgroup are measured simultaneously using only one circuit (hereafter, such subgroups are called "partitions"). Notably, the partitioning method is also useful for simulating the time evolution of Hamiltonians in terms of reducing algorithmic errors induced via Trotter decomposition [19] and quantum phase–estimation algorithm with ancilla qubits [20]. To maximize the effectiveness of simultaneous measurement, the number of partitions should be minimized. However, minimizing the number of partitions is an NP-hard problem. To date, several algorithms have been proposed to address such problems [10-14]. They are mainly based on either the maximum clique searching method [11,14,21] or the graph coloring method [11-14]. For the maximum clique searching method, two algorithms are mainly used: Boppana–Halldórsson algorithm [22] and Bron–Kerbosch algorithm [23,24]. Their time complexity and solution optimality, however, have a trade-off relation. For the Boppana–Halldórsson algorithm, the time complexity along the number of Pauli strings is a polynomial but the optimality of solutions is not guaranteed. The Bron–Kerbosch algorithm can guarantee the optimality but exhibits exponential time complexity. Regarding the graph coloring method, Verteletskyi et al. [11,12] and Hamamura et al. [13] tested the performance of the largest-first method, which has been proven to afford the best performance among various heuristic orderings. This method takes the polynomial time with the number of Pauli strings but provides less optimal partitioning results than maximum clique searching algorithms with the Bron–Kerbosch algorithm [14] when the number of Pauli strings exceeds ~500. An algorithm-specific partitioning method proposes a partitioning scheme using the nature of the qubit-mapping methods of Hamiltonians in quantum chemistry [14]. This partitioning method may not require long time, but the resultant reduction factor is only 8 at maximum (when the Jordan–Wigner method is used).

Such classical algorithms with polynomial time scaling can reduce the number of measurements for estimating the expectation values of Hamiltonians to some extent. However, we need to further reduce the number of measurements as far as possible for the following reason. As mentioned above, the estimation of expectation value of the given Hamiltonian is subject to classical optimization (i.e., the expectation value has to be measured per optimization step) and the number of the optimization steps sharply increases with increasing number of parameters. Therefore, the factor of how the number of measurements is reduced using the partitioning method must be evaluated. However, although considerable effort has been made to improve the reduction factor, it cannot reach the factor obtained using the Bron–Kerbosch algorithm (which requires the exponential time). Therefore, we believe that an application-specific computer that operates on a different calculation principle from a conventional computer would be necessary for drastic improvement in both time complexity and the resultant reduction factor.

Another method to reduce the total number of measurements is the shadowing method, which is based on classical shadowing [15,16]. The advantage of this method is that in some cases, a fewer number of required measurements can be realized compared with the partitioning method by efficiently determining the basis-changing operation per measurement [16]. However, in the worst case, the method requires the same number of quantum circuits as that of the measurements, which may become an additional and non-negligible cost for hardware experiment. Moreover, to determine per-measurement basis-changing operations, the computational cost of $\Omega(n_{\text{meas}})$ is required, where $n_{\text{meas}}$ denotes the total number of measurements. Yen et al. [17] discussed a combination of the shadowing and partitioning methods to further reduce the total number of measurements. From this point of view, it would also be of great help if the computational time for determining the necessary quantum circuits and the number of such circuits for the partitioning method could be considerably reduced.

In this paper, we propose a fast, effective, and versatile algorithm to address such partitioning problems with Ising machines. The proposed algorithm is based on the maximum clique searching method, and we transform maximum clique searching problems into quadratic unconstrained binary optimization (QUBO) problems, which are equivalent to the Ising model optimization problems [25]. This approach allows us to use Ising machines to quickly solve the problems. Using Fujitsu's second-generation Digital Annealer (DA), a hardware architecture designed to efficiently solve QUBO problems [26], as an Ising machine, we demonstrate that the performance of our algorithm is much better in terms of time complexity and solution optimality by comparing it with existing algorithms (Boppana–Halldórsson algorithm and Bron–Kerbosch algorithm). In addition to the partitioning problem for VQE Hamiltonians, we tested our algorithm on a full tomography of one $n$-qubit quantum state (where the number of Pauli strings to be measured is $4^n - 1$) to benchmark the results to the theoretical ones [27]. Our algorithm can be applied to the problems larger than the capacity of an Ising machine by using it repeatedly, as will be shown below.

This paper is organized as follows. Section 2 describes the theoretical background and general procedure of performing simultaneous measurements. Section 3 explains our new Ising model–based partitioning algorithm and how the DA works. Section 4 describes the performance of the new algorithm in measuring the expectation values of multiple Pauli strings, comparing it with existing maximum clique searching algorithms. Section 5 summarizes this study and discusses future perspectives.

## 2. Simultaneous Measurements

In this section, we explain the technical background of simultaneous measurements and partitioning. In Subsection 2.1, we describe the relation between the commutativity of Pauli strings and simultaneous measurements. In Subsection 2.2, we present the general procedure of simultaneous measurements.

### 2.1. Theoretical background

A simultaneous measurement is based on the fact that the expectation values of two Pauli strings $P_1$ and $P_2$ can be simultaneously estimated by applying an appropriate basis-changing operation if and only if they commute each other (i.e., $P_1 P_2 = P_2 P_1$) [10]. Generally, given a quantum state $\rho$, the expectation value of an observable $M$ is $\mathrm{Tr}(M\rho)$. When the matrix $B$ can diagonalize the matrix $M$ so that $Z_{\{i_j\}} = BMB^{-1}$, this condition leads to

$$\mathrm{Tr}(M\rho) = \mathrm{Tr}\left[Z_{\{i_j\}}(B\rho B^{-1})\right], \tag{5}$$

where $Z_{\{i_j\}}$ denotes the tensor matrix of $Z$ and $I$:

$$Z_{\{i_j\}} = \bigotimes_{j=1}^{n} Z^{i_j}, \tag{6}$$

$$Z = \begin{bmatrix} 1 & 0 \\ 0 & -1 \end{bmatrix}, I = \begin{bmatrix} 1 & 0 \\ 0 & 1 \end{bmatrix}, \tag{7}$$

where $j$ denotes the qubit index beginning from 1 and $i_j \in \{0,1\}$ for each $j$. Thus, we can estimate $\mathrm{Tr}(M\rho)$ by applying $B$ to $\rho$ as a basis-changing operation and performing a projective measurement of the qubits labeled with the sequence $\{j | i_j = 1\}$ along the computational basis ($Z$-basis).

Here, we suppose that $\rho$ is an $n$-qubit state and $B$ is a basis-changing operation. For any of $2^n - 1$ possible $\{i_j\}$ (excluding $i_j = 0$ for all the $j$ values), $M_{\{i_j\}}$ exists, which satisfies

$$Z_{\{i_j\}} = BM_{\{i_j\}}B^{-1}. \tag{8}$$

As shown in eq 5, eq 8 satisfies

$$\mathrm{Tr}\left(M_{\{i_j\}}\rho\right) = \mathrm{Tr}\left[Z_{\{i_j\}}(B\rho B^{-1})\right]. \tag{9}$$

From eq 8, the commutation relation between $M_{\{i_j\}}$ and $M_{\{i_j\}'}$ can be derived as

$$M_{\{i_j\}}M_{\{i_j\}'} = B^{-1}Z_{\{i_j\}}Z_{\{i_j\}'}B = B^{-1}Z_{\{i_j\}'}Z_{\{i_j\}}B$$
$$= M_{\{i_j\}'}M_{\{i_j\}}, \tag{10}$$

for any $\{i_j\}, \{i_j\}'$. Therefore, for $M_{\{i_j\}}$ and $M_{\{i_j\}'}$ commute for any $\{i_j\}, \{i_j\}'$ is the requirement for simultaneous measurements. Moreover, for any partition $\{M_{\{i_j\}}\}$, the existence of a basis-changing gate $B$ that satisfies eq 9 has been proven in a previous study [10].

The commutativity of Pauli strings has two settings [14]. One is qubit-wise commutativity (QWC), which means that for every qubit, corresponding Pauli operators commute each other. The other is general commutativity (GC), which means that Pauli strings commute as a whole, while each Pauli operator does not necessarily commute. In the GC setting, for some qubits, corresponding Pauli operators can be anticommuting. Two Pauli strings commute when the number of the anticommuting pairs of Pauli operators is even. When we apply QWC, the basis-changing gate $B$ can be described as a tensor product of a single-qubit gate. Meanwhile, the resultant number of partitions is $3^n$ for an $n$-qubit full tomography (each of the partitions contains one Pauli string described as a tensor product of $X$, $Y$, and $Z$). For estimating the expectation values of Hamiltonians for VQE, a previous study [11] shows that the number of partitions is only three times less than the number of Pauli strings. By contrast, when we apply GC, $B$ is described as entangled gates and the number of partitions is expected to be less than that for QWC. For $n$-qubit full tomography, when GC is applied, $4^n - 1$ Pauli strings can be divided into $2^n + 1$ partitions, each of which contains $2^n - 1$ Pauli strings [27]. In this study, we used GC to investigate the maximum effect of simultaneous measurements.

### 2.2. General procedure and its time complexity

Figure 1 shows a scheme for measuring the expectation values of multiple Pauli strings using simultaneous measurement and partitioning. After enumerating the Pauli strings that are required to estimate the expectation values (step (1)), the commutativity of each pair of Pauli strings was checked (step (2)). Then, we created partitions where all Pauli strings commute (step (3)). Based on this, we determined the basis-changing gate B for each partition (step (4)), and finally, all the expectation values of the Pauli strings were estimated (step (5)). Using steps (2) and (3) presented in Figure 1, the number of circuits was reduced to the number of partitions.

To evaluate the overall performance of partitioning, the time complexity and solution optimality should be examined. The time complexity of step (2) is $O(N^2 n)$, where $N$ denotes the number of Pauli strings and $n$ denotes the number of qubits. Conversely, the time complexity of step (3) strongly depends on the algorithm used for the partitioning.

Two partitioning algorithms are generally used for solving the maximum clique searching problem. The Boppana–Halldórsson algorithm [22] uses a greedy method for creating each partition, which does not necessarily result in a maximum-size partition. It has a roughly quadratic time complexity with no guarantee of optimality, although its worst-case time complexity is not well studied. Meanwhile, the Bron–Kerbosch algorithm [23,24] uses a rigorous method for creating each partition to obtain a maximum-size partition. It has an exponential time complexity of $O(3^{n/3})$ for creating one partition as the worst case [24] but yields an optimal solution. More details of these algorithms are described in Appendix 1.

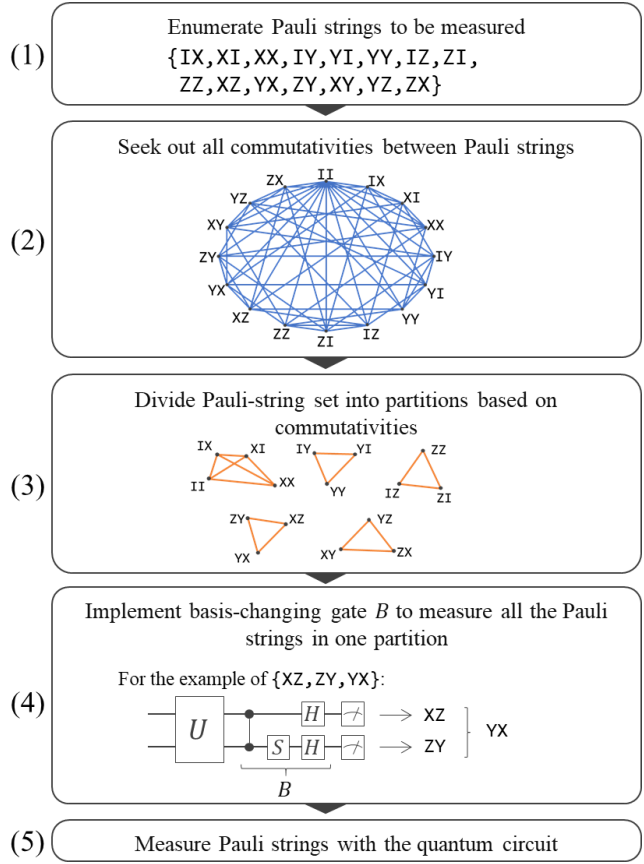

**Figure 1**. Schematic of measuring the expectation values of multiple Pauli strings using simultaneous measurements

## 3. Methods

In this section, we describe our proposed Pauli string partitioning methods. Subsection 3.1 introduces the proposed Ising model–based partitioning algorithm. Subsections 3.2 and 3.3 describe the specific partitioning problems we address in this study and the settings of DA, respectively.

### 3.1. Ising model–based partitioning algorithm

The proposed algorithm is based on maximum clique searching [21]. A partition with the maximum number of elements is created using Pauli strings, and the process is repeated with other Pauli strings until no other string remains, as shown in Figure 2. In this algorithm, we break down each partition-creating problem into a QUBO problem, which is equivalent to an Ising model problem [25] and can be solved efficiently using an Ising machine. In this QUBO problem, each Pauli string is assigned a binary variable, and the value of the variables distinguishes whether the corresponding Pauli strings are included in a target partition.

Suppose that we create one maximum-size partition from the candidate Pauli string group $\{P_1, \cdots, P_N\}$. To map this problem to a QUBO problem, the cost function should be determined so that it is minimized when the number of Pauli strings in the target partition is maximized. The cost function can be defined as follows:

$$f(x_1, \ldots, x_N) = -\sum_{1 \leq i \leq N} b_i x_i + \sum_{\substack{1 \leq i \leq N \\ 1 \leq j \leq N}} m c_{i,j} x_i x_j, \tag{11}$$

where $x_1, \ldots, x_N \in \{0,1\}$ denotes the binary variables mapped to Pauli strings. $x_k = 1$ means that the Pauli string $P_k$ is included in a target partition; $x_k = 0$ means otherwise. $b_1, \ldots, b_N$ denote positive constants. $c_{1,1}, \ldots, c_{N,N}$ denote nonnegative constants, satisfying $c_{i,j} = 0$ if $P_i P_j = P_j P_i$; otherwise, $c_{i,j} = 1$. $m$ denotes a positive constant. In the right hand of eq 11, the first term means that the cost function decreases as the number of Pauli strings in the target clique increases. The second term means that if pairs of Pauli strings that do not commute exist in a partition, then the cost function increases. The value of $m$ must be selected to satisfy the condition that the contribution of the second term is zero when the global minimum of eq 11 is realized. $m > 1$ satisfies this condition for any case, and we set $m = 2$ in this study. When the cost function reaches the global minimum, the obtained subgroup of Pauli strings $\{P_i | x_i = 1\}$ can be considered the partition that has the largest number of elements in a Pauli string set of interest. The pedagogical example of this Ising model mapping is presented in Figure 3.

When the number of variables $\{x_i\}$ is less than or equal to the number of variables that an Ising machine can handle ($n_{\text{bit}}$), we can directly determine $\{x_i\}$, which minimizes the target cost function using an Ising machine [26,28,29]. Ising machines are designed to solve such Ising-type optimization problems by setting the initial values of $\{x_i\}$ first, finding more optimal values that decrease the target cost function, and updating them iteratively. Although the computational principle is different in each problem, Ising machines are generally designed to solve these problems faster than conventional computers.

When the number of Pauli strings exceeds $n_{\text{bit}}$, additional procedures are required. Given the Pauli string group $\mathcal{P} = \{P_1, \ldots, P_{n_{\text{bit}}}, \ldots, P_N\}$, the partition $\mathcal{C}$ is created using the following procedures in our Ising model–based algorithm, as shown in Figure 4. (1) A subgroup $\mathcal{S}_\mathcal{P}$ that comprises the first $n_{\text{bit}}$ elements of $\mathcal{P}$ is defined, and a partition $\mathcal{D} = \{P_{d_1}, \ldots, P_{d_p}\}$ is created using $\mathcal{S}_\mathcal{P}$ by solving the corresponding QUBO (eq 11) problem with an Ising machine. (2) $o_\mathcal{D}(P_j)$ is defined for each $P_j \in \mathcal{P}$, which denotes the number of Pauli strings $P_i \in \mathcal{D}$ that satisfy $P_i P_j = P_j P_i$, and $\mathcal{P} = \{P_j\}$ is sorted in a descending order of $o_\mathcal{D}(P_j)$. (3) A subgroup $\mathcal{T}_\mathcal{P} \subset \mathcal{P}$ is defined, such that $\mathcal{T}_\mathcal{P}$ comprises the first $n_{\text{bit}}$ elements of the sorted $\{P_j\}$. (4) A partition $\mathcal{C} = \{P_{c_1}, \ldots, P_{c_q}\}$ is created using $\mathcal{T}_\mathcal{P}$ by solving the corresponding QUBO (eq. 11) problem with the Ising machine. In this case, the Ising machine is used twice (in creating $\mathcal{D}$ and then $\mathcal{C}$) for a single cycle of the partitioning process. Procedures from (2) to (4) can be repeated $r$ times to create an optimal partition, with $\mathcal{D} \leftarrow \mathcal{C}$ being updated between procedures (2) and (4). In this study, we set $r = 1$. The flowchart of this algorithm is presented in Figure 5, and the overall partitioning algorithm is shown in Algorithm 1.

We investigated the performance of this Ising model–based algorithm in terms of time complexity and solution optimality compared with those of the Boppana–Halldórsson and Bron–Kerbosch algorithms, both of which are also based on maximum clique searching. For the Boppana–Halldórsson and Bron–Kerbosch algorithms, we used NetworkX [30] implemented in Python because both algorithms are implemented in it and it allows us to benchmark them easily.

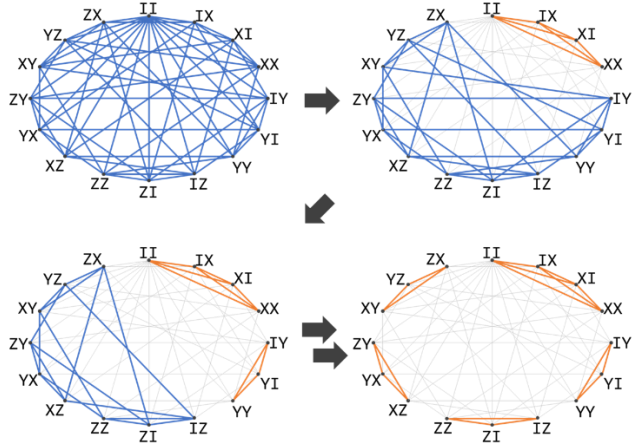

**Figure 2**. Illustration of the partitioning process for two-qubit full tomography. An edge between two nodes denotes that the corresponding two Pauli strings commute. One partition is described as the node groups, with thick edges colored in orange. In this example, the 16 Pauli strings can be divided into five groups.

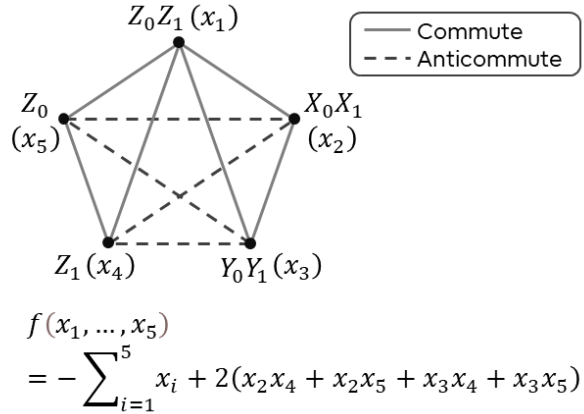

**Figure 3**. Illustration of the mapping of partitioning problems to the Ising model equation (eq 8), where we set $m = 2$. Here, we are creating a maximum-size partition from $\{Z_0Z_1, X_0X_1, Y_0Y_1, Z_1, Z_0\}$. The result of this equation is minimized when $\{x_1, x_2, x_3, x_4, x_5\} = \{1,1,1,0,0\}, \{1,0,0,1,1\}$, which reflect the maximum-size partitions.

### 3.2. Overview of DA and its settings

In this study, we used DA as an Ising machine for the following reasons. First, all the variables in DA are fully connected, which is preferable for the problems under consideration because for any variable $x_i$, roughly $N/2$ variables $x_j$ exist that satisfy $c_{i,j} \neq 0$. Second, DA can rapidly search the $2^{n_{\text{bit}}}$ space (where $n_{\text{bit}}$ denotes the number of variables in DA) to obtain a (globally) minimum value of QUBO problems, such as eq. (11). It is because of an efficient parallel trial scheme for a Markov chain Monte Carlo method combined with massive parallelization and a dynamic escaping function from local minima [31]. DA can generally solve such QUBO problems much faster than simulated annealing conducted on a classical computer [31].

All calculations using DA were conducted under a computational environment prepared for research use. The second-generation DA [26] that we used herein allows us to tune the available number of variables up to 8,192. In this study, without further notice, all calculations using DA were conducted with $n_{\text{bit}} = 8{,}192$. For each calculation, the number of Monte Carlo steps was fixed to $10^8$.

---

**Algorithm 1**: Ising model–based algorithm

---

1: **Input**: a set of Pauli strings $\mathcal{P} = \{P_1, \cdots, P_N\}$, interaction coefficients $\{c_{i,j} | 1 \leq i \leq N, 1 \leq j \leq N\}$ ($c_{i,j} = 0$ if $P_i P_j = P_j P_i$, $c_{i,j} = 1$ otherwise), coefficients $\{b_i | 1 \leq i \leq N\}$ ($b_i = 1$ for any $i$), available number of variables of Ising machine $n_{\text{bit}}$, positive coefficient $m$, positive integer $r$

2:     Set: $\mathcal{P}_1 = \{P_1, \cdots, P_N\}$

3:     Set: $k = 1$

4:     **while** $\mathcal{P}_k \neq \emptyset$ **do**

5:       **if** $|\mathcal{P}_k| > n_{\text{bit}}$ **then**

6:         Sort $P_j \in \mathcal{P}_k$ in ascending order of index $j$

7:         Set: $\mathcal{S}_{\mathcal{P}_k} \subset \mathcal{P}_k$ such that $\mathcal{S}_{\mathcal{P}_k}$ consists of the first $n_{\text{bit}}$ elements of the sorted $\mathcal{P}_k$

8:         Determine binary variables $\{x_i | P_i \in \mathcal{S}_{\mathcal{P}_k}\}$ by Ising machine, which minimize eq 11 subject to $x_i = 0$ for all of $\{x_i | P_i \notin \mathcal{S}_{\mathcal{P}_k}\}$

9:         Set: $\mathcal{D}_k = \{P_i | P_i \in \mathcal{S}_{\mathcal{P}_k}, x_i = 1\}$

10:         **for** $l = 1, \cdots, r$ **do**

11:           Set: $o_{\mathcal{D}_k}(P_j) = |\{P_i | c_{i,j} = 0, P_i \in \mathcal{D}_k\}|$ for each $P_j \in \mathcal{P}_k$

12:           Sort $P_j \in \mathcal{P}_k$ in descending order of $o_{\mathcal{D}_k}(P_j)$

13:           Set: $\mathcal{T}_{\mathcal{P}_k} \subset \mathcal{P}_k$ such that $\mathcal{T}_{\mathcal{P}_k}$ consists of the first $n_{\text{bit}}$ elements of the sorted $\mathcal{P}_k$

14:           Determine binary variables $\{x_i | P_i \in \mathcal{T}_{\mathcal{P}_k}\}$ by Ising machine, which minimize eq 11 subject to $x_i = 0$ for all of $\{x_i | P_i \notin \mathcal{T}_{\mathcal{P}_k}\}$

15:           $\mathcal{D}_k \leftarrow \{P_i | P_i \in \mathcal{T}_{\mathcal{P}_k}, x_i = 1\}$

16:           **if** $l = r$ **then**

17:             Set: $\mathcal{C}_k = \mathcal{D}_k$

18:           **end if**

19:         **end for**

20:       **else**

21:         Determine binary variables $x_1, \cdots, x_n$ which minimize eq 11 by Ising machine

22:         Set: $\mathcal{C}_k = \{P_i | x_i = 1\}$

23:       **end if**

24:       Set: $\mathcal{P}_{k+1} = \mathcal{P}_k - \mathcal{C}_k$

25:       $k \leftarrow k + 1$

26:     **end while**

27: **Output**: $\mathcal{C}_1, \mathcal{C}_2, \cdots$

---

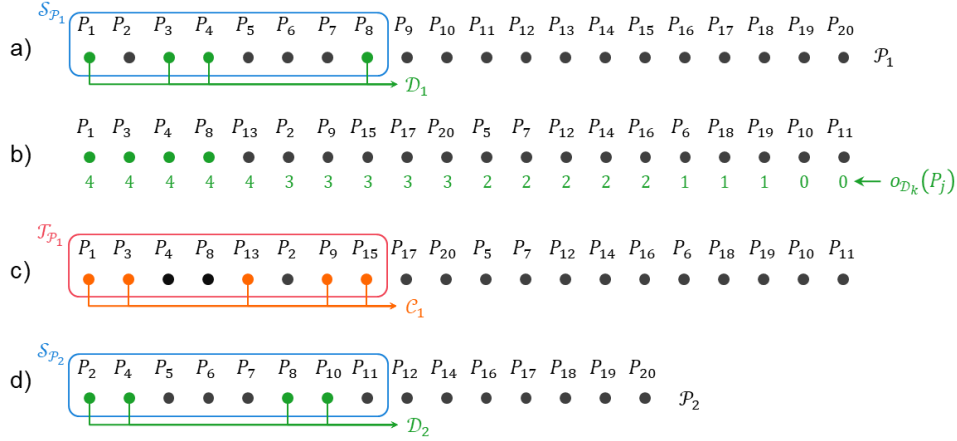

**Figure 4.** Illustration of the partitioning procedure when $N > n_{\text{bit}}$ ($N = 20$ and $n_{\text{bit}} = 8$). a) Determine $\mathcal{S}_{\mathcal{P}_1}$ as the first $n_{\text{bit}}$ elements of $\mathcal{P}_1$ and calculate $\mathcal{D}_1$ using DA. b) Calculate $o_{\mathcal{D}_1}(P_j)$ for each $P_j \in \mathcal{P}_1$ and then sort $\mathcal{P}_1$ in a descending order of $o_{\mathcal{D}_1}(P_j)$. c) Determine $\mathcal{T}_{\mathcal{P}_1}$ as the first $n_{\text{bit}}$ elements of the sorted $\mathcal{P}_1$ and calculate $\mathcal{C}_1$ using DA. d) Determine $\mathcal{P}_2 = \mathcal{P}_1 - \mathcal{C}_1$ and sort it in an ascending order of index $j$. Then, continue partitioning until $\mathcal{P}_k = \emptyset$.

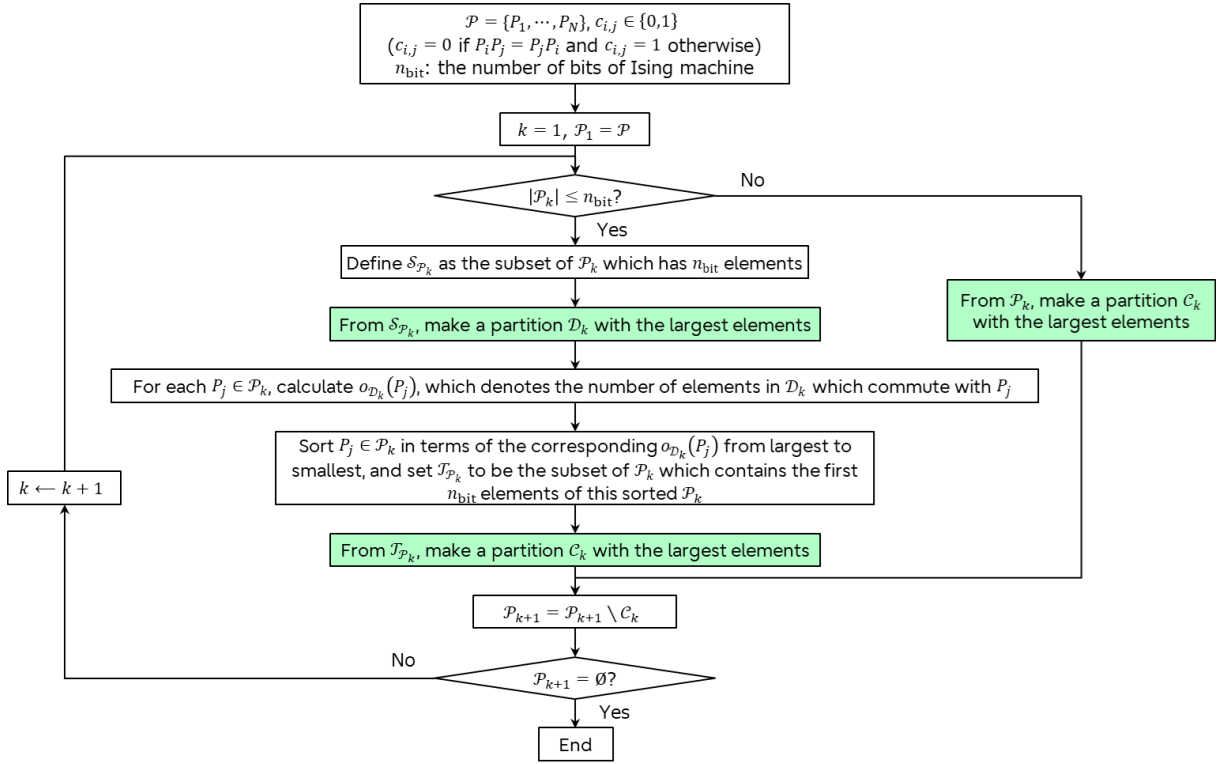

**Figure 5.** Flowchart of partitioning with the Ising model–based algorithm. Ising machines are employed in the green-colored steps.

### 3.3. Partitioning problems

In this study, we set the partitioning problems of $n$-qubit full tomography ($n = 1, \ldots, 8$) and estimate the expectation values of Hamiltonians for VQE. For the VQE problems, we set the target molecules as H2, LiH, H2O, and CH4 in the STO-3G basis set and BeH2, H2O, N2, and NH3 in the 6-31G basis set. To create corresponding Hamiltonians, we set their molecular configurations by referring to [11] and used the Jordan–Wigner qubit-mapping method. For each molecule in the STO-3G basis set, we assumed a variable number of spatial orbitals in an active space, as summarized in Table 1. For each partitioning problem, we excluded the identity Pauli string $\otimes_{j=1}^{n} I$ because it commutes with all other Pauli strings and its expectation value is always 1.

For the full-tomography Pauli string set, we performed indexing of Pauli strings in the ascending order of $\sum_{i=1}^{n} p^{(i)} 4^{n+1-i}$, where $p^{(i)} = 0$ if the Pauli operator of $i$th qubit is $I$ and, similarly, 1 if $X$, 2 if $Y$, and 3 if $Z$. For estimating the expectation values of Hamiltonians for VQE, we performed indexing of Pauli strings along with the OpenFermion ordering [11,32].

**Table 1**. Pauli string set for VQE (excluding identity)

| molecules | basis set | qubit mapping | # of spatial orbitals | # of qubits | # of Pauli strings |
|---|---|---|---|---|---|
| H2 | STO-3G | Jordan–Wigner | 1 | 2 | 3 |
| H2 | STO-3G | Jordan–Wigner | 2 | 4 | 14 |
| LiH | STO-3G | Jordan–Wigner | 3 | 6 | 117 |
| LiH | STO-3G | Jordan–Wigner | 4 | 8 | 192 |
| LiH | STO-3G | Jordan–Wigner | 5 | 10 | 275 |
| LiH | STO-3G | Jordan–Wigner | 6 | 12 | 630 |
| H2O | STO-3G | Jordan–Wigner | 4 | 8 | 220 |
| H2O | STO-3G | Jordan–Wigner | 5 | 10 | 311 |
| H2O | STO-3G | Jordan–Wigner | 6 | 12 | 740 |
| H2O | STO-3G | Jordan–Wigner | 7 | 14 | 1,389 |
| CH4 | STO-3G | Jordan–Wigner | 4 | 8 | 240 |
| CH4 | STO-3G | Jordan–Wigner | 5 | 10 | 591 |
| CH4 | STO-3G | Jordan–Wigner | 6 | 12 | 1,518 |
| CH4 | STO-3G | Jordan–Wigner | 7 | 14 | 3,005 |
| CH4 | STO-3G | Jordan–Wigner | 8 | 16 | 5,236 |
| CH4 | STO-3G | Jordan–Wigner | 9 | 18 | 8,479 |
| BeH2 | 6-31G | Jordan–Wigner | 13 | 26 | 9,203 |
| H2O | 6-31G | Jordan–Wigner | 13 | 26 | 12,731 |
| N2 | 6-31G | Jordan–Wigner | 18 | 36 | 34,622 |
| NH3 | 6-31G | Jordan–Wigner | 15 | 30 | 52,805 |

### 4. Results and Discussions
#### 4.1. Time complexity and solution optimality when $N \leq n_{\text{bit}}$

In this section, we discuss the time complexity and solution optimality for Pauli string partitioning problems when the number of Pauli strings $N$ subject to partitioning does not exceed $n_{\text{bit}}$. Figure 4 shows the plots of algorithm runtime $t$ against the number of Pauli strings $N$ using the three algorithms, with their regression curves in the form of $t = aN^b$. The time complexity on $N$ for the Ising model–based algorithm within the range of $N \leq n_{\text{bit}} = 8,192$ was estimated as $N^{0.52}$. This condition contrasts with $N^{2.57}$ for the Boppana–Halldórsson algorithm and $N^{5.41}$ for the Bron–Kerbosch algorithm. These results suggest that the Ising model–based algorithm is the most scalable algorithm among the three in terms of time complexity. As shown in Figure 6, with the processing ability of our laptop resources, the Ising model–based algorithm showed better performance than the Bron–Kerbosch algorithm when $N \geq 300$ and the Boppana–Halldórsson algorithm when $N \geq 2{,}000$. Although the runtime performances of the Boppana–Halldórsson and Bron–Kerbosch algorithms depend on how much computer resources one can utilize, the threshold number of Pauli strings $N_{\text{th}}$ exists, such that the Ising model–based algorithm shows the best performance when $N > N_{\text{th}}$.

We investigated the performance guarantee for the time complexity for the Ising model–based algorithm. Each Monte Carlo step in a DA calculation takes the same amount of time [31]. Thus, the time for creating one partition is constant in our setting, where the number of Monte Carlo steps per DA calculation is constant. Figure 7 shows that the overall runtime $t$ of Algorithm 1 depends almost linearly on the resultant number of partitions. Here, creating one partition takes 4 to 5 s. With the introduction of $\tau$, which denotes the average runtime for one DA calculation, the Ising model–based algorithm runtime is written as

$$t = \frac{\tau N}{F}, \qquad (12)$$

where $F$ denotes the reduction factor [14], which is the value defined as the number of Pauli strings $N$ divided by the resultant number of partitions. The observed dependence $t \propto N^{0.52}$ reflects that the dependence of $F$ on $N$ is $F \propto N^{0.48}$. However, it also reflects the specific characteristic of partitioning problems, i.e., the Pauli strings for VQE observables and full tomography. This dependence of $F$ cannot apply to all types of partitioning problems. The performance guarantee can be investigated by assuming that $F = 1$ as the worst case. With its application to eq. 12, the worst-case time complexity can be confirmed as $O(N)$, which still has a lower dimension than those of the Bron–Kerbosch and Boppana–Halldórsson algorithms.

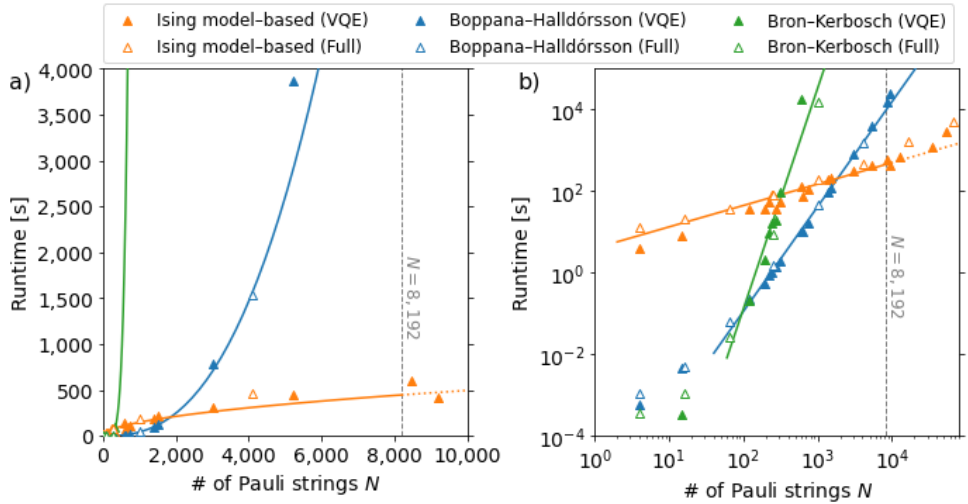

**Figure 6**. Plot of the runtime along the number of observables obtained using the Ising model–based algorithm (orange), Boppana–Halldórsson algorithm (blue), and Bron–Kerbosch algorithm (green). (a) Plotted in the linear axis, (b) plotted in the logarithmic axis. The regression curve for each algorithm is in the $t = aN^b$ form. The regression curve for the Ising model–based algorithm is determined based on the data plots for $N < 8{,}192$.

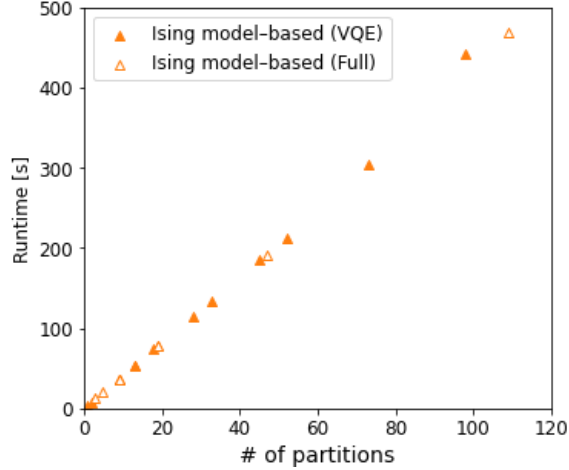

**Figure 7**. Plot of the partitioning runtime along the resultant number of partitions using the Ising model–based algorithm

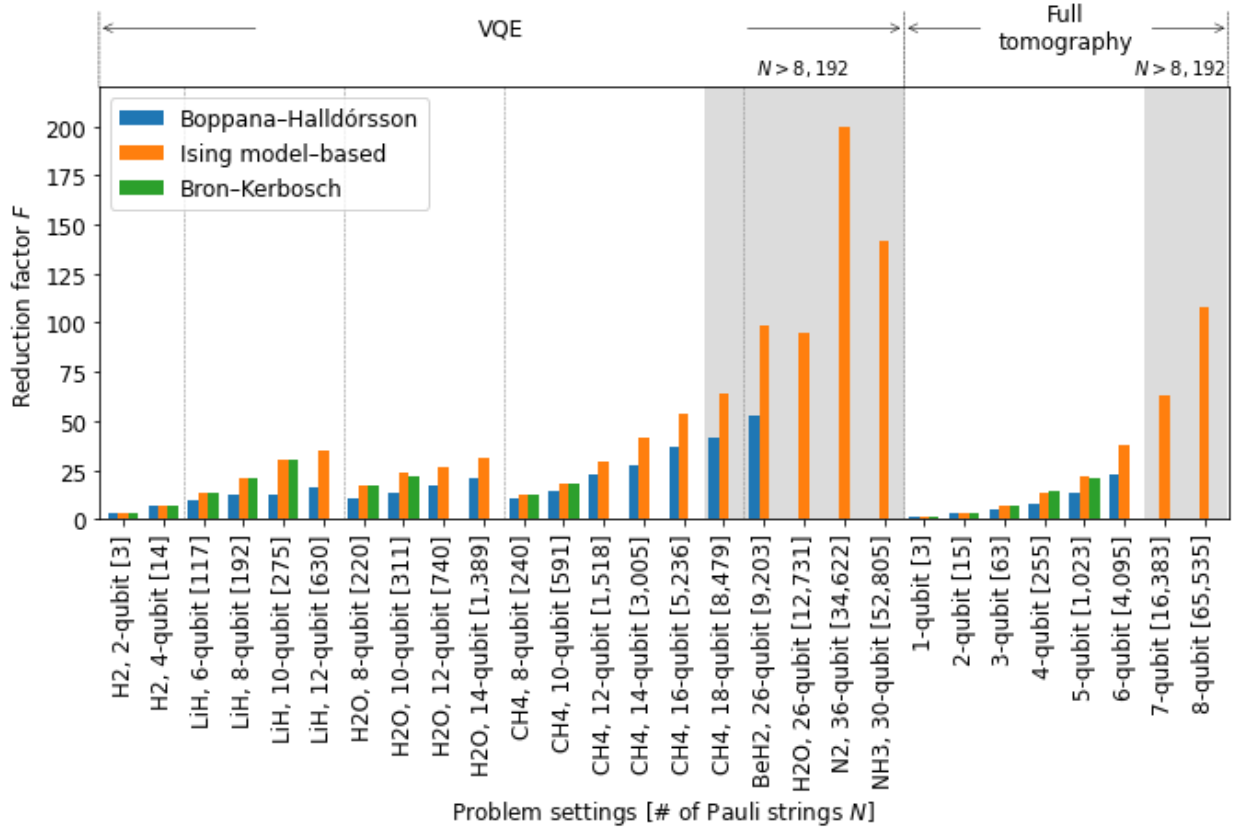

**Figure 8**. Bar plot of the reduction factor $F$ (defined in the main text) using the Ising model–based algorithm (orange), Boppana–Halldórsson algorithm (blue), and Bron–Kerbosch algorithm (green). For the Ising model–based algorithm, we set $n_{\text{bit}} = 8{,}192$. For the Boppana–Halldórsson and Bron–Kerbosch algorithms, some results are not shown because they are assumed to take more than 10 h.

Next, we discuss the solution optimality of each algorithm. We evaluated it using the reduction factor $F$; a large reduction factor means a high solution optimality. Figure 8 shows the reduction factor for each partitioning problem. When the number of Pauli strings is small ($< 20$), the reduction factors are similar among the algorithms. By contrast, when the number of Pauli strings increases, the resultant reduction factors of the Ising model–based algorithm and Bron–Kerbosch algorithm become larger than that of the Boppana–Halldórsson algorithm. This result obtained because of the characteristics of the three algorithms. The Ising model–based algorithm and Bron–Kerbosch algorithm maximize the number of elements of a partition-per-partition creation process, whereas the Boppana–Halldórsson algorithm does not always maximize the number of elements of one partition because it incorporates the greedy approach. This point can be observed in Figure 9; for the Boppana–Halldórsson algorithm, the number of elements of each partition is smaller than that for the other algorithms. Notably, for $n$-qubit full tomography for $n \geq 4$, the resultant reduction factor $F$ is less than $2^n + 1$, which is proven to be the maximum value [26]. The results show that the maximum reduction factor has not yet been obtained even via our method. This is probably because a partition that is chosen in the one partition-creating process is not always the ideal one that realizes the maximum effect. Each partition-creating process affects the subsequent partition-creating processes, and the remaining Pauli string groups may often have only partitions with less than $2^n - 1$ Pauli strings.

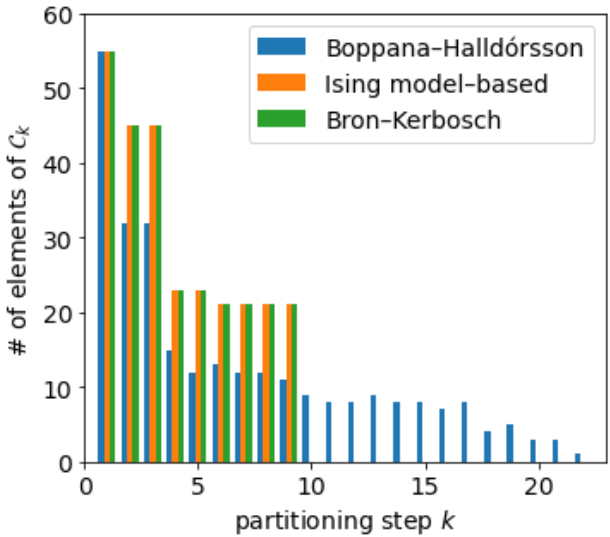

**Figure 9**. Number of Pauli strings of each partition as a result of partitioning 275 Pauli strings (LiH as a basis set of STO-3G and five spatial orbitals in the active space)

### 4.2 Time complexity and solution optimality when $N > n_{\text{bit}}$

Section 4.1 presents the excellent partitioning performance of the Ising model–based algorithm for $N \leq n_{\text{bit}}$. However, in quantum chemical calculations, $N$ may exceed $n_{\text{bit}}$. In this section, we examine the performance of the Ising model–based algorithm when $N > n_{\text{bit}}$.

We discuss the solution optimality first. We have shown the performance of the Ising model–based algorithm when $n_{\text{bit}} = 8{,}192$ and $N > 8{,}192$ in Figures 6 and 8. The runtimes required using the Ising model–based algorithm are within 1.5 h and remain much shorter than those required in the Boppana–Halldórsson algorithm even when $N > 8{,}192$ (Figure 4), and a much better solution optimality (i.e., larger reduction factor $F$) is realized. We demonstrated that $F \approx$ 200 when solving the partitioning problem of $N = 34{,}622$ (Figure 6). This result strongly suggests that the extension of the Ising model–based algorithm to the cases of $N > n_{\text{bit}}$ is effective.

However, solution optimality and time complexity for the case of $N > n_{\text{bit}}$ are anticipated to be quantitatively different from those for the case of $N \leq n_{\text{bit}}$. In the following subsection, we will discuss their differences in detail.

### 4.2.1 Solution optimality

When $N > n_{\text{bit}}$, a natural assumption is that the performance of the Ising model–based algorithm degrades as $N$ increases with respect to $n_{\text{bit}}$. To confirm this, we compared the Ising model–based algorithm performance in the cases of $n_{\text{bit}} = 8{,}192$ and $n_{\text{bit}} = 1{,}024$, as shown in Table 2. The number of partitions is larger when $n_{\text{bit}} = 1{,}024$, which leads to the decrease in the reduction factor $F$. This finding suggests that the size of $n_{\text{bit}}$ indeed affects the performance.

Figure 8 shows the number of elements of each partition $C_k$ ($k = 1, \cdots, 100$) for some of the partitioning results for $N > 8{,}192$ with the conditions of $n_{\text{bit}} = 8{,}192$ and $n_{\text{bit}} = 1{,}024$. For $n_{\text{bit}} = 1{,}024$, the resultant number of elements of each partition $C_k$ is substantially smaller than that for $n_{\text{bit}} = 8{,}192$, especially when a partition with >200 elements. This condition can be a direct cause of the increasing number of partitions. However, executing two DA calculations (one to determine $\mathcal{D}_k$ and another to determine $C_k$) is more effective than executing one DA calculation to determine $\mathcal{D}_k$ and then regarding it as $C_k$ because the number of elements of $C_k$ is larger than that of $\mathcal{D}_k$ for each $k$ (Figure 10). Furthermore, the ratio of the number of elements of $C_k$ to that of $\mathcal{D}_k$ is greater in the case of $n_{\text{bit}} = 1{,}024$ than in the case of $n_{\text{bit}} = 8{,}192$ (Figure 11). This finding suggests that the effect of two DA calculations is greater when the size of $N$ with respect to $n_{\text{bit}}$ is larger.

To further investigate how the relation between $N$ and $n_{\text{bit}}$ influences its performance, a convenient step is to introduce a parameter $D$, which denotes the relative dimension of the partitioning problem against $n_{\text{bit}}$. We define $D$ as

$$D \equiv \begin{cases} N/n_{\text{bit}} & (\text{if } N > n_{\text{bit}}) \\ 1 & (\text{if } N \leq n_{\text{bit}}) \end{cases}. \tag{13}$$

We first investigated the performance of the Ising model–based algorithm for the partitioning problems of $N > 8{,}192$ with various $D$ (by varying $n_{\text{bit}} \in \{8{,}192, 4{,}096, 2{,}048, 1{,}536, 1{,}024, 768, 512, 384, 256, 192\}$ on solving the same partitioning problem). Figure 10b shows the logarithmic plots of the resultant reduction factor $F$ along $D$. The reduction factor $F$ slightly decreases with increasing $D$ for all the $N$ values. Therefore, we can expect that the reduction factor is maximum under the $D = 1$ ($N \leq n_{\text{bit}}$) condition.

Then, we investigated the performance of the Ising model–based algorithm when $N < 8{,}192$ with $D \geq 1$, including $D = 1$. For these calculations, we set $n_{bit} \in \{8{,}192, 4{,}096, 2{,}048, 1{,}536, 1{,}024, 512, 256, 192, 128, 96, 64, 48\}$. Figure 12a shows the plots of the resultant reduction factors $F_D$ as a function of $D$. The reduction factor $F_D$ decreases as $D$ increases, but its ratio to $F_1$, i.e.,

$$p_D \equiv \frac{F_D}{F_1}, \tag{14}$$

is not less than 0.9 until $D \approx 10$. As $d$ increases above ~10, we observed that $p_D$ substantially decreases. When $D$ is sufficiently large, the dependence of $p_D$ on $d$ is denoted by $O(D^{-1})$ (see Appendix 2 in Supplementary Information).

These results show that although the Ising model–based algorithm is effective even when $N > n_{\text{bit}}$, the larger $n_{\text{bit}}$ provides better optimality for partitioning the problems with $N > n_{\text{bit}}$.

**4.2.2 Time complexity**

Next, we discuss the time complexity. As described in Figure 6b, the partitioning runtime $t$ for $N > 8{,}192$ shows some deviations from the extrapolated regression curve ($t = aN^{0.52}$), which is determined from the runtime data for $N \leq 8{,}192$. When $N > n_{\text{bit}}$, the partitioning runtimes would be longer owing to the following factors: (i) a greater number of resultant partitions leads to more DA calculation step and (ii) two DA calculations are required to create a single partition as long as the number of remaining Pauli strings exceeds $n_{\text{bit}}$. The algorithm runtime $t_D$ for a relative dimension $D$ is described as

$$t_D = \frac{\tau_D (1+s) N}{F_D}, \tag{15}$$

where $s$ denotes the ratio of the number of partitions whose creations require two DA calculations to that of all partitions and $\tau_D$ denotes the average time for one DA calculation. From eqs 12 and 14, $t_D$ can be rewritten as

$$t_D = \frac{\tau_D}{\tau_1}(1+s) p_D^{-1} t_1. \tag{16}$$

As described in eq. 16, three contributions ($\tau_D/\tau_1$, $1+s$, and $p_D^{-1}$) are made to the deviation of $t_D$ from $t_1$. Among them, within the range of our simulation ($N \leq 65{,}535$), the difference between the actual and estimated runtimes from the extrapolated regression curve (in the case of $n_{\text{bit}} = 8{,}192$) is mainly contributed by $1+s$, i.e., up to 1.6. $p_D^{-1}$, which is expected to be $\leq 1.1$ because $D \leq 8$ and $\tau_D/\tau_1$ is $\sim 1$. However, when a partitioning problem with an even larger number of Pauli strings $N$ is assumed to be solved, the contribution of $1+s$ is $O(1)$ because $s \leq 1$ by definition. Similarly, $\tau_D/\tau_1 = O(1)$. The contribution of $p_D^{-1}$ becomes the main factor of time complexity, which evolves as $O(N)$ (Figure 13 and Appendix 2). Therefore, with the assumption of $t_{D=1} \propto N^{0.52}$ even when $N > 8{,}192$, the time complexity for $N > 8{,}192$ is estimated as $N^{1.52}$.

Next, we discuss the worst-case time complexity. Even when assuming $t_1 = O(N)$ as the worst case, we can confirm that $t_D = O(N^2)$, which still has a lower complexity than the observed time complexity of the Boppana–Halldórsson algorithm. Moreover, when the overall partitioning procedure depicted in Figure 1 is considered, a notable detail is observed that if a sufficiently large $n_{\text{bit}}$ is available for large $N$, then the partitioning procedure (3) may not be the most rate-limiting step in the overall procedure of measuring the expectation values of multiple Pauli strings (Figure 1) because the commutativity evaluation step (2) of Pauli strings (Figure 1) has a higher dimension of the time complexity of $O(N^2 n)$.

We can consider several additional strategies to improve the partitioning method for future studies. For example, setting different $\mathcal{S}_{\mathcal{P}_k}$ by sorting $\mathcal{P}_k$ along different orders [11] potentially results in a more optimal $\mathcal{T}_{\mathcal{P}_k}$ than along the OpenFermion order, for obtaining $\mathcal{C}_k$ with more elements. Moreover, repeating the partitioning steps from (2) to (4) $r$ times can increase the number of elements of $\mathcal{C}_k$. However, the number of elements in a partition cannot exceed $n_{\text{bit}}$ when using the Ising model–based algorithm, which would be the theoretical limit of the algorithm. In addition, enabling a larger $n_{\text{bit}}$ than 8,192 with the future development of annealing machines would ensure a shorter algorithm runtime by reducing the contributions of $1 + s$ and $p_D^{-1}$, whereas it potentially increases $t_1$ by increasing $\tau_1$.

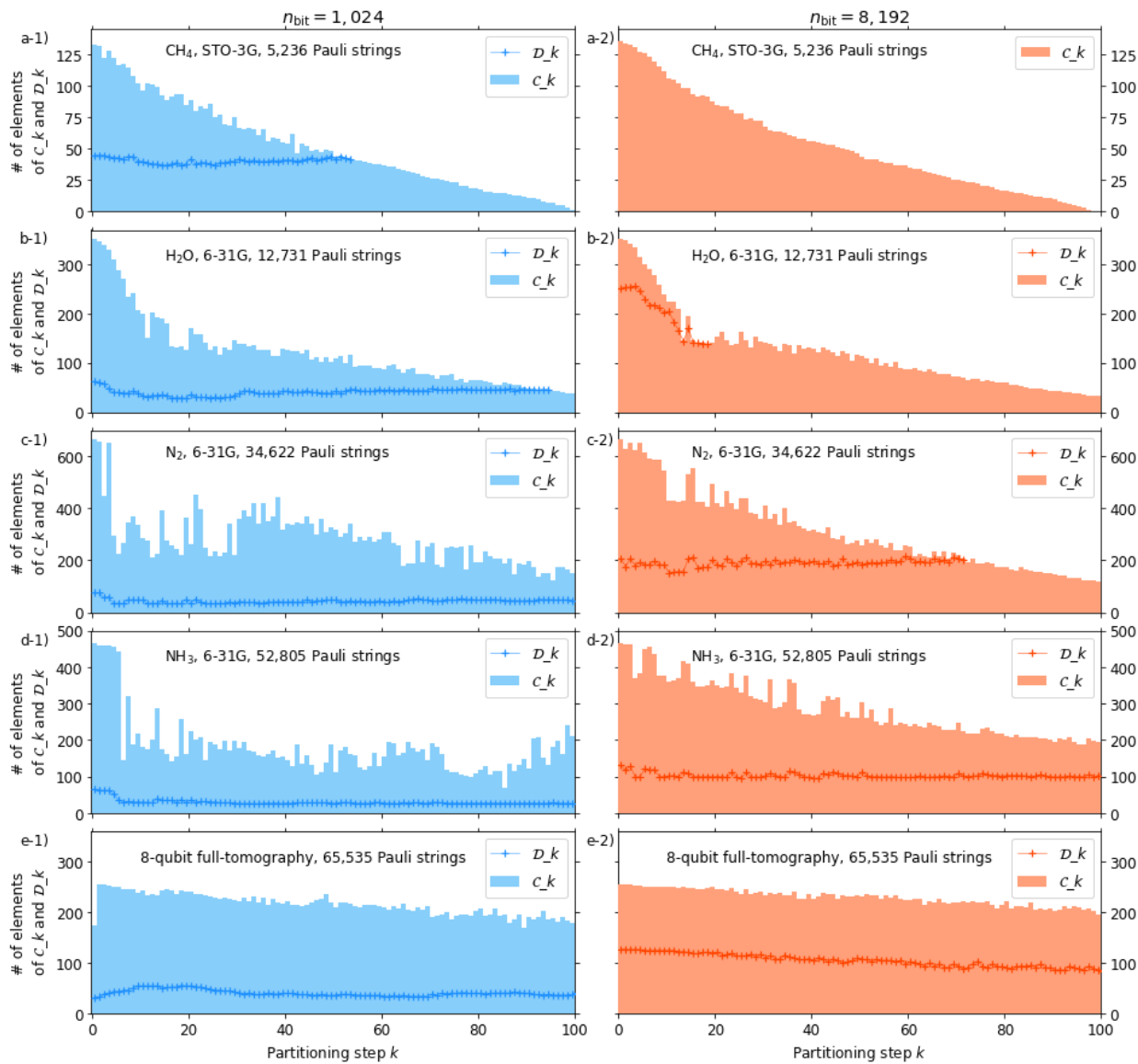

**Figure 10**. Bar plots of the number of elements of $\mathcal{C}_k$ and the number of elements of $\mathcal{D}_k$ (as defined in Algorithm 1) per partitioning step $k$ (in a range of $1 \leq k \leq 100$) for partitioning problems of (a) $CH_4$ in STO-3G basis set (8 spatial orbitals), (b) $H_2O$ in 6-31G basis set (13 spatial orbitals), (c) $N_2$ in 6-31G basis set (18 spatial orbitals), (d) $NH_3$ in 6-31G basis set (15 spatial orbitals), and (e) eight-qubit full tomography.

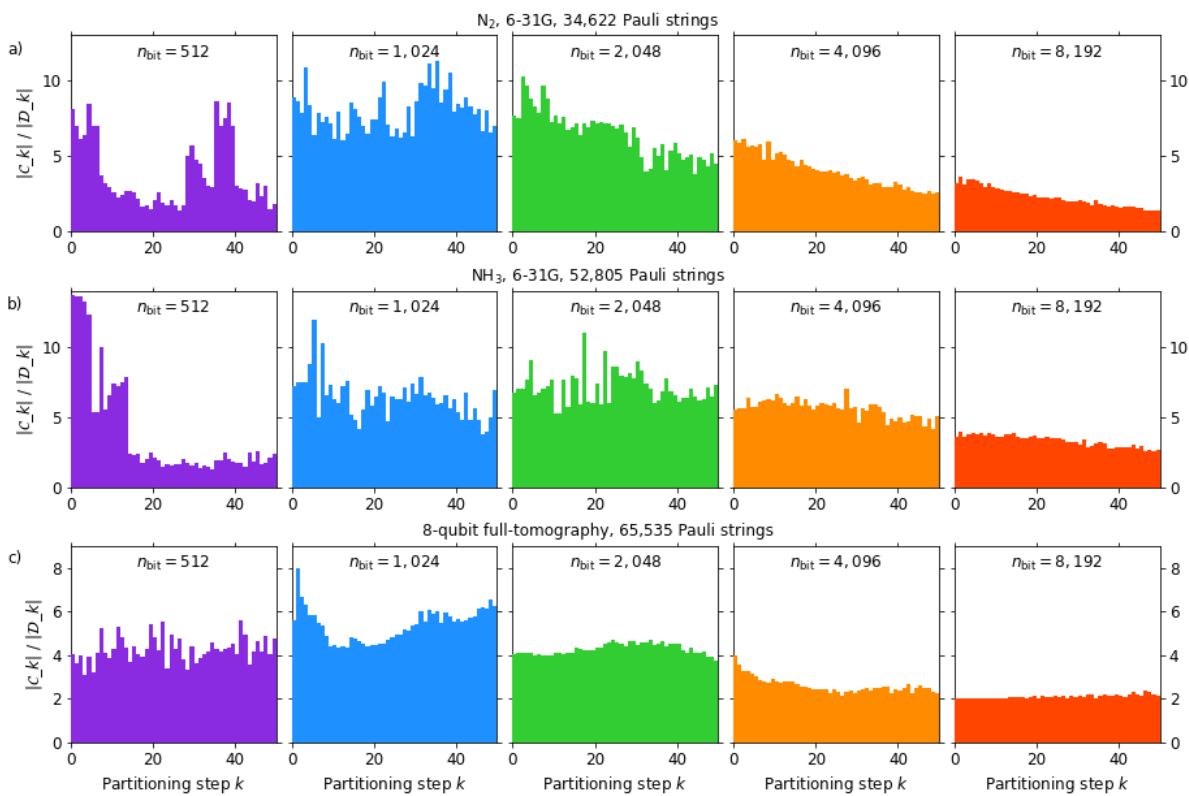

**Figure 11**. Bar plots of the number of elements of $\mathcal{C}_k$ divided by that of $\mathcal{D}_k$ (as defined in Algorithm 1) per partitioning step $k$ (in a range of $1 \leq k \leq 50$) for partitioning problems of (a) $N_2$ in the 6-31G basis set (18 spatial orbitals), (b) $NH_3$ in the 6-31G basis set (15 spatial orbitals), and (c) eight-qubit full tomography.

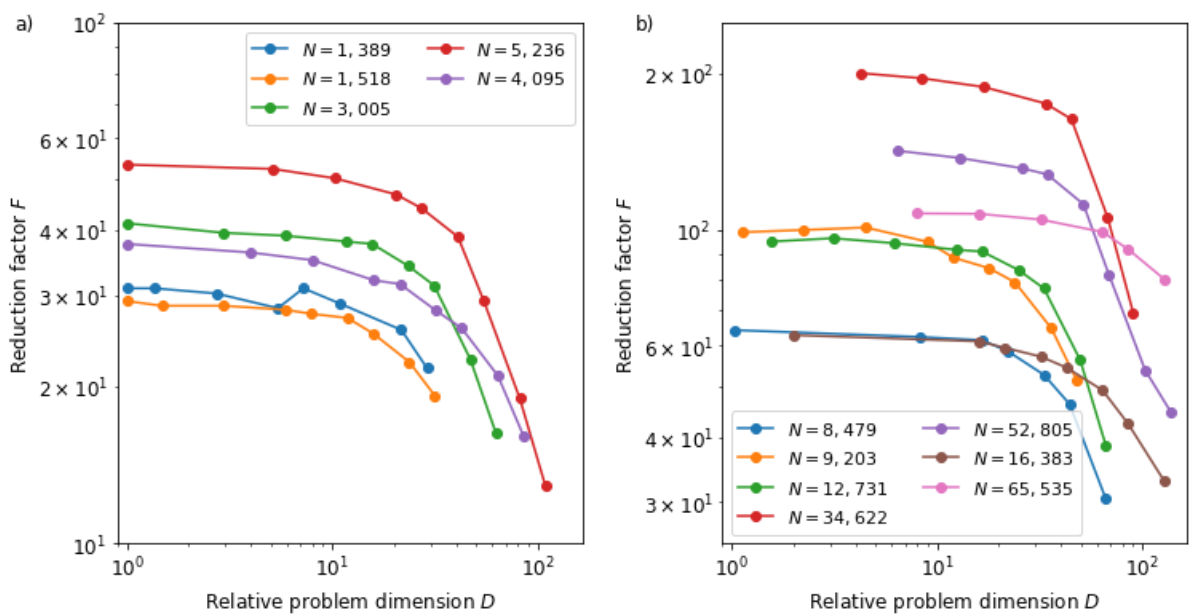

**Figure 12**. Plots of the reduction factor $F$ as a function of the relative problem dimension $D$ for the partitioning problems of a) $N \leq 8{,}192$ and b) $N > 8{,}192$

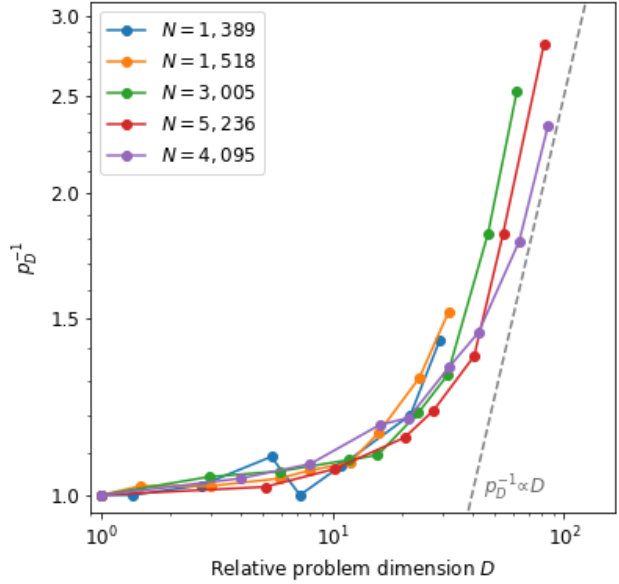

**Figure 13**. Plots of $p_D^{-1}$ as a function of the relative problem dimension $D$ for the partitioning problems of $N \leq 8{,}192$

**Table 2**. Partitioning performance of the Ising model–based algorithm ($n_{\text{bit}} = 8{,}192$ mode and $n_{\text{bit}} = 1{,}024$ mode) and Boppana–Halldórsson algorithm in the case of $N > 8{,}192$. For each problem and each IM algorithm mode, the relative problem dimension $D$ (defined in the main text), number of resultant partitions, reduction factor, number of DA calculations, and calculation time are shown.

| | | | | | | Ising model–based algorithm ($n_{\text{bit}} = 8{,}192$) | | | | | Ising model–based algorithm ($n_{\text{bit}} = 1{,}024$) | | | | | Boppana–Halldórsson algorithm | | |
|---|---|---|---|---|---|---|---|---|---|---|---|---|---|---|---|---|---|---|
| Problem type | molecule | # of spatial orbitals | basis set | # of qubits | # of Pauli strings | $D$ | # of partitions | Reduction factor | # of DA calculations | calculation time | $D$ | # of Partitions | Reduction factor | # of DA calculations | calculation time | # of Partitions | Reduction factor | calculation time |
| VQE | $CH_4$ | 9 | STO-3G | 18 | 8479 | 1.04 | 132 | 64.2 | 134 | 598.7 | 8.28 | 136 | 62.3 | 222 | 903.6 | 203 | 41.8 | 15795.2 |
| VQE | $BeH_2$ | 13 | 6-31G | 26 | 9203 | 1.12 | 93 | 99.0 | 96 | 421.0 | 8.99 | 97 | 94.9 | 156 | 637.1 | 176 | 52.3 | 24525.9 |
| VQE | $H_2O$ | 13 | 6-31G | 26 | 12731 | 1.56 | 134 | 95.0 | 153 | 711.3 | 12.43 | 139 | 91.6 | 234 | 953.4 | N/A | N/A | N/A |
| VQE | $N_2$ | 18 | 6-31G | 36 | 34622 | 4.23 | 173 | 200.1 | 245 | 1153.3 | 33.81 | 198 | 174.9 | 356 | 1450.2 | N/A | N/A | N/A |
| VQE | $NH_3$ | 15 | 6-31G | 30 | 52805 | 6.45 | 372 | 141.9 | 586 | 2915.0 | 51.57 | 471 | 112.1 | 892 | 3635.9 | N/A | N/A | N/A |
| Full tomography | - | - | - | 7 | 16383 | 2.00 | 261 | 62.8 | 338 | 1601.0 | 16.00 | 268 | 61.1 | 470 | 1906.0 | N/A | N/A | N/A |
| Full tomography | - | - | - | 8 | 65535 | 8.00 | 609 | 107.6 | 985 | 5027.1 | 64.00 | 661 | 99.1 | 1246 | 5073.7 | N/A | N/A | N/A |

## 5. Conclusions and Future Perspective

Herein, we propose a method for the partitioning of Pauli strings. We transfer the portioning to Ising optimization and solve it using an Ising machine. Compared with conventional algorithms (the Boppana–Halldórsson and Bron–Kerbosch algorithms), our Ising model–based algorithm shows advantages in terms of solution optimality and time complexity as the number of Pauli strings increases. Therefore, we believe the proposed method to be one of the most useful methods for solving the partitioning problem, especially when large quantum systems become available in the future. Moreover, the proposed method is versatile for any sort of quantum algorithms.

To effectively perform such simultaneous measurements, the implementation of basis-changing gate sets $B$ (as denoted in step (4) in Figure 1) is necessary. When GC is applied for partitioning, constituting $B$ is known to be a nontrivial problem [14]. In addition, such gate sets $B$ include two-qubit entangling gates (e.g., CNOT), which is likely to be noisier than single-qubit gates. In this context, the reduction of the number of two-qubit gates in $B$ is an important problem. A universal method for constituting $B$ is available in Ref. [14]. Another strategy is to tune the partitioning process to suppress the number of entangling gates [13]. To make this strategy compatible with the Ising model–based algorithm, we may need to tune the coefficients $c_{i,j}$ to circumvent too many entangling gates, which will be the subject of future works.

**Appendix 1:** Overviews of the Boppana–Halldórsson and Bron–Kerbosch algorithms

Here, we show the detailed descriptions of the Boppana–Halldórsson algorithm (Algorithm 2) and Bron–Kerbosch algorithm (Algorithm 3). They are both based on the maximum clique searching method.

A. Boppana–Halldórsson algorithm

In the Boppana–Halldórsson algorithm [22], a partition $\mathcal{C}_k$ is created based on a greedy method. Following this algorithm, a partition $\mathcal{C}(\mathcal{X})$ is created from a set of Pauli strings $\mathcal{X}$ with the following steps. The overall description of this algorithm is shown in Algorithm 2.

(i) Randomly choose Pauli string $X_1 \in \mathcal{X}$ and define $\mathcal{S}_{\mathcal{C}(X_1)}(\mathcal{X})$ as the subset of $\mathcal{X}$, which contains all the Pauli strings (other than $X_1$) that commute with $X_1$, and define $\mathcal{S}_{\mathcal{A}(X_1)}(\mathcal{X})$ as the subset of $\mathcal{X}$, which contains all the Pauli strings that anticommute with $X_1$.

(ii) If $\mathcal{S}_{\mathcal{C}(X_1)}(\mathcal{X})$ is not empty, randomly choose $X_2$ from $\mathcal{S}_{\mathcal{C}(X_1)}(\mathcal{X})$ and then define $\mathcal{S}_{\mathcal{C}(X_1),\mathcal{C}(X_2)}(\mathcal{X})$ as the subset of $\mathcal{S}_{\mathcal{C}(X_1)}(\mathcal{X})$, which contains all the Pauli strings (other than $X_2$) that commute with $X_2$, and define $\mathcal{S}_{\mathcal{C}(X_1),\mathcal{A}(X_2)}(\mathcal{X})$ as the subset of Pauli strings, all of which anticommute with $X_2$.

(iii) Similar to (ii), if $\mathcal{S}_{\mathcal{A}(X_1)}(\mathcal{X})$ is not empty, randomly choose $X_3$ from $\mathcal{S}_{\mathcal{A}(X_1)}(\mathcal{X})$. Then, define $\mathcal{S}_{\mathcal{A}(X_1),\mathcal{C}(X_2)}(\mathcal{X})$ as the subset of $\mathcal{S}_{\mathcal{A}(X_1)}(\mathcal{X})$, which contains all the Pauli strings (other than $X_3$) that commute with $X_3$, and define $\mathcal{S}_{\mathcal{A}(X_1),\mathcal{A}(X_2)}(\mathcal{X})$ as the subset of Pauli strings, all of which anticommute with $X_3$.

(iv) Divide $\mathcal{X}$ into subsets $\mathcal{S}_{Q_1,\cdots,Q_m}(\mathcal{X})$, where $Q_k \in \{\mathcal{C}(X_k), \mathcal{A}(X_k)\}$ for $1 \leq k \leq m$, in the similar manner as mentioned in the aforementioned procedure until all the $\mathcal{S}_{Q_1,\cdots,Q_m}(\mathcal{X})$ become empty.

(v) With such divided subgroups, define the subsets $\mathcal{C}\left(\mathcal{S}_{Q_1,\cdots,Q_k}(\mathcal{X})\right)$ for the descending manner of $k$. First, $\mathcal{C}\left(\mathcal{S}_{Q_1,\cdots,Q_k}(\mathcal{X})\right)$ is determined to be empty when $\mathcal{S}_{Q_1,\cdots,Q_k}(\mathcal{X})$ is empty. When $\mathcal{S}_{Q_1,\cdots,Q_k}(\mathcal{X})$ is not empty, determine $\mathcal{C}\left(\mathcal{S}_{Q_1,\cdots,Q_k}(\mathcal{X})\right)$ to be a larger subset among $\{X_{k+1}\} \cup \mathcal{C}\left(\mathcal{S}_{Q_1,\cdots Q_k,\mathcal{C}(X_{k+1})}(\mathcal{X})\right)$ and $\mathcal{C}\left(\mathcal{S}_{Q_1,\cdots Q_k,\mathcal{A}(X_{k+1})}(\mathcal{X})\right)$ after determining $\mathcal{C}\left(\mathcal{S}_{Q_1,\cdots,Q_k,\mathcal{C}(X_{k+1})}(\mathcal{X})\right)$ and $\mathcal{C}\left(\mathcal{S}_{Q_1,\cdots,Q_k,\mathcal{A}(X_{k+1})}(\mathcal{X})\right)$. Notably, all the elements commute one another, irrespective of the chosen subset.

(vi) Finally, $\mathcal{C}(\mathcal{X})$ is determined as a larger subset among $\{X_1\} \cup \mathcal{S}_{\mathcal{C}(X_1)}(\mathcal{X})$ and $\mathcal{S}_{\mathcal{A}(X_1)}(\mathcal{X})$.

**Algorithm 2**: Boppana–Halldórsson algorithm

1: **Input**: a set of Pauli strings $\mathcal{P} = \{P_1, \cdots, P_N\}$, interaction coefficients $\{c_{i,j} | 1 \leq i \leq N, 1 \leq j \leq N\}$ ($c_{i,j} = 0$ if $P_i P_j = P_j P_i$, $c_{i,j} = 1$ otherwise)
2: **function** Ramsey($\mathcal{X}$)
3:    **if** $\mathcal{X} = \emptyset$ **then**
4:       **return** $\emptyset$
5:    **else**
6:       Randomly choose $X = P_i$ from the elements of $\mathcal{X}$
7:       **Set**: $\mathcal{S}_{\mathcal{C}(X)}(\mathcal{X}) \equiv \{P_j | P_j \in \mathcal{X}, j \neq i, c_{j,i} = 0\}$
8:       **Set**: $\mathcal{S}_{\mathcal{A}(X)}(\mathcal{X}) \equiv \{P_j | P_j \in \mathcal{X}, j \neq i, c_{j,i} = 1\}$
9:       **return** larger of $\{P_i\}$ + Ramsey($\mathcal{S}_{\mathcal{C}(X)}(\mathcal{X})$), Ramsey($\mathcal{S}_{\mathcal{A}(X)}(\mathcal{X})$)
10:   **end if**
11: **Set**: $k = 1$
12: **Set**: $\mathcal{P}_1 = \mathcal{P}$
13: **while** $\mathcal{P}_k \neq \emptyset$ **do**
14:    **Set**: $\mathcal{C}_k = $ Ramsey($\mathcal{P}_k$)
15:    **Set**: $\mathcal{P}_{k+1} = \mathcal{P}_k - \mathcal{C}_k$
16:    $k \leftarrow k + 1$
17: **end while**
18: **Output**: $\mathcal{C}_1, \mathcal{C}_2, \cdots$

## B. Bron–Kerbosch algorithm

In the Bron–Kerbosch algorithm [23,24], many partition candidates $\{c_k(\mathcal{X})\}$ are created and, among these candidates, a partition with the maximum number of elements is chosen. We show the steps of creating $\{c_k(\mathcal{X})\}$ below and in Algorithm 3:

(i) Let $k = 1$.

(ii) Let $c_k(\mathcal{X})$ be empty; then, randomly choose a Pauli string $X_1 \in \mathcal{X}$, and add it to $c_k(\mathcal{X})$.

(iii) Define $N_{X_1}(\mathcal{X})$ as the subset of $\mathcal{X}$, where all the elements commute with $X_1$; then, randomly choose $X_2$ from $N_{X_1}(\mathcal{X})$, and add it to $c_k(\mathcal{X})$.

(iv) Define $N_{X_2}(\mathcal{X})$ as the subset of $\mathcal{X}$, where all the elements commute with $X_2$. Then, randomly choose $X_3$ from $N_{X_1}(\mathcal{X}) \cap N_{X_1}(\mathcal{X})$, and add it to $c_k(\mathcal{X})$.

(v) Repeat step (iv) until $N_{X_1}(\mathcal{X}) \cap \cdots \cap N_{X_m}(\mathcal{X})$ becomes empty, and finally let $c_k(\mathcal{X}) = \{X_1, \cdots, X_m\}$.

(vi) Let $k \leftarrow k + 1$, and repeat steps (ii)–(v) with different Pauli strings $X_1, X_2, \cdots$.

(vii) Repeat step (vi) for all considerable Pauli strings $X_1, X_2, \cdots$.

The corresponding module in NetworkX [30] incorporates protocols to efficiently narrow down the candidate partitions $\{c_k(\mathcal{X})\}$ [24].

**Algorithm 3**: Bron–Kerbosch algorithm

1: **Input**: a set of Pauli strings $\mathcal{P} = \{P_1, \cdots, P_n\}$, interaction coefficients $\{c_{i,j} | 1 \leq i \leq N, 1 \leq j \leq N\}$ ($c_{i,j} = 0$ if $P_i P_j = P_j P_i$, $c_{i,j} = 1$ otherwise)
2: **function** Expand($d, l_1, \mathcal{P}', \mathcal{N}, \mathcal{F}$)
3:     Set: $l = l_1$
4:     Choose $P_m \in \mathcal{N}$ which maximize the number of $P_{m'} \in \mathcal{N}$ satisfying $c_{m,m'} = 0$ and $m \neq m'$
5:     Set: $\mathcal{N}' = \{P_{m'} \in \mathcal{N} | c_{m',m} = 1 \text{ or } m' = m\}$
6:     **for** $P_i \in \mathcal{N}' \cap \mathcal{F}$ **do**
7:         Set: $d' = d + \{P_i\}$
8:         Set: $N_{P_i} = \{P_j \in \mathcal{P}' | j \neq i, c_{j,i} = 0\}$
9:         $\mathcal{F} \leftarrow \mathcal{F} + \{P_i\}$
10:        **if** $N_P \cap \mathcal{N} = \emptyset$ **then**
11:            $c_l(\mathcal{X}) = d'$
12:            $l \leftarrow l + 1$
13:        **else**
14:            $\{c_l(\mathcal{X}), \cdots, c_{l'}(\mathcal{X})\}, l' = $ Expand$(d', l, \mathcal{P}', N_{P_i} \cap \mathcal{N}, \mathcal{F})$
15:            $l \leftarrow l' + 1$
16:        **end if**
17:     **end for**
18:     **return** $\{c_{l_1}(\mathcal{X}), \cdots, c_{l-1}(\mathcal{X})\}, l - 1$
19: Set: $k = 1$
20: Set: $\mathcal{P}_1 = \mathcal{P}$
21: **while** $\mathcal{P}_k \neq \emptyset$ **do**
22:     Set: $l_0 = 0$
23:     Set: $\mathcal{F} = \emptyset$
24:     **for** $P_i \in \mathcal{P}_k$ **do**
25:         Set: $d = \emptyset$
26:         $l_0 \leftarrow l_0 + 1$
27:         $d \leftarrow d + \{P_1\}$
28:         $\mathcal{F} \leftarrow \mathcal{F} + \{P_1\}$
29:         Set: $N_{P_i} = \{P_j \in \mathcal{P}_k | i \neq j, c_{i,j} = 0\}$
30:         $\{c_{l_0}, \cdots, c_l\}, l = $ SunFunc$(d, l_0, \mathcal{P}_k, N_{P_i}, \mathcal{F})$
31:         $l_0 \leftarrow l$
32:     Set $C_k$ to be the set with the largest elements among $\{c_1(\mathcal{X}), \cdots, c_{l_0}(\mathcal{X})\}$
33:     Set: $\mathcal{P}_{k+1} = \mathcal{P}_k - C_k$
34:     $k \leftarrow k + 1$
35: **end while**

**Appendix 2:** Proof that the $D$ dependence of $p_D$ is represented by $O(D^{-1})$

When the Ising model–based algorithm is used, the maximum number of elements in each resultant partition is $n_{\text{bit}}$. Therefore, the reduction factor $F_D$ does not exceed $n_{\text{bit}}$. Here, we define a factor $\gamma_D \leq 1$, which satisfies

$$\gamma_D = \frac{F_D}{n_{\text{bit}}}. \tag{A1}$$

Therefore, $\gamma_D = 1$ if all the resultant partitions contain $n_{\text{bit}}$ elements; otherwise $\gamma_D < 1$. When $n_{\text{bit}}$ is close to 1 (i.e., $D$ is close to $N$), a reasonable assumption is that each partition that contains $C_k$ also contains $n_{\text{bit}}$ (or very close to $n_{\text{bit}}$) elements. Thus, the factor $\gamma_D$ increases and becomes close to 1 (Figure A1). Therefore, by applying this to eq 14, we obtain

$$p_D^{-1} = \frac{F_1}{F_D} = \frac{F_1}{\gamma_D} n_{\text{bit}}^{-1} \geq F_1 n_{\text{bit}}^{-1} = \frac{F_1 D}{N} \tag{A2}$$

and

$$p_D^{-1} = \frac{F_1}{F_D} = \frac{\gamma_1 N}{\gamma_D n_{\text{bit}}} \leq \frac{N}{n_{\text{bit}}} = D. \tag{A3}$$

Therefore, we can confirm that $p_D^{-1} = O(D)$ (Figure 13).

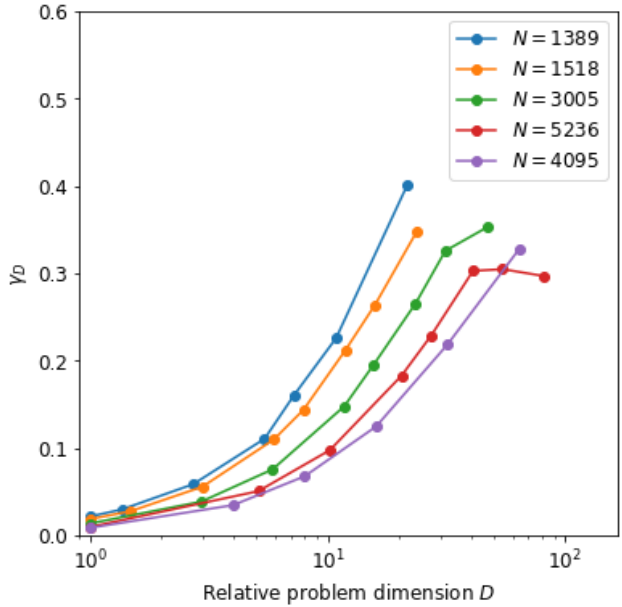

Fig. A1. Plots of $\gamma_D$ along the relative problem dimension $D$

ACKNOWLEDGMENT

The authors thank Yoshinori Tomita, Toshiyuki Miyazawa, and Kazuya Takemoto for support for the utilization of DA and fruitful discussions.